\documentclass[10pt,journal,compsoc]{IEEEtran}
\usepackage{ragged2e}

\ifCLASSOPTIONcompsoc
  \usepackage[nocompress]{cite}
\else
  \usepackage{cite}
\fi
\ifCLASSINFOpdf
\else
\fi

\usepackage{graphicx}
\usepackage{amsmath,amssymb} % define this before the line numbering.
\usepackage{color}
\usepackage[linesnumbered,ruled]{algorithm2e}

\usepackage{caption}
\usepackage{subcaption}
\usepackage{booktabs}
\usepackage{amssymb}
\usepackage{tabularx}
\newcolumntype{C}[1]{>{\centering\arraybackslash}p{#1}}

\newif\ifshowcomments
\showcommentstrue
\def\ie{\emph{i.e.}}
\def\eg{\emph{e.g.}}
\def\etal{{\em et al.}}
\def\etc{{\em etc.}}
\newcommand{\para}[1]{\vspace{.05in}\noindent\textbf{#1}}

\begin{document}
%\title{Bare Demo of IEEEtran.cls for\\ IEEE Computer Society Journals}
%
\title{DNF-Net: a Deep Normal Filtering Network\\ for Mesh Denoising}


\author{Xianzhi Li,
        Ruihui Li,
        Lei Zhu,
	Chi-Wing~Fu,~\IEEEmembership{Member,~IEEE,}
	and~Pheng-Ann~Heng,~\IEEEmembership{Senior Member,~IEEE}% <-this % stops a space
\IEEEcompsocitemizethanks{\IEEEcompsocthanksitem X. Li, R. Li, L. Zhu, C.-W. Fu, and P.-A. Heng are with the Chinese University of Hong Kong. P.-A. Heng is also with Guangdong-Hong Kong-Macao Joint Laboratory of Human-Machine Intelligence-Synergy Systems, Shenzhen Institutes of Advanced Technology, Chinese Academy of Sciences, Shenzhen 518055, China.\protect\\
E-mail: \{xzli, lirh, lzhu, cwfu, pheng\}@cse.cuhk.edu.hk
}% <-this % stops an unwanted space
}


\markboth{Submitted to IEEE Transactions on Visualization and Computer Graphics}
{(under review)}
%{Shell \MakeLowercase{\textit{et al.}}: Bare Demo of IEEEtran.cls for Computer Society Journals}


\IEEEspecialpapernotice{
	\begin{center}
		\renewcommand\thefigure{\arabic{figure}}
		\setcounter{figure}{0}
		\includegraphics[width=0.99\textwidth]{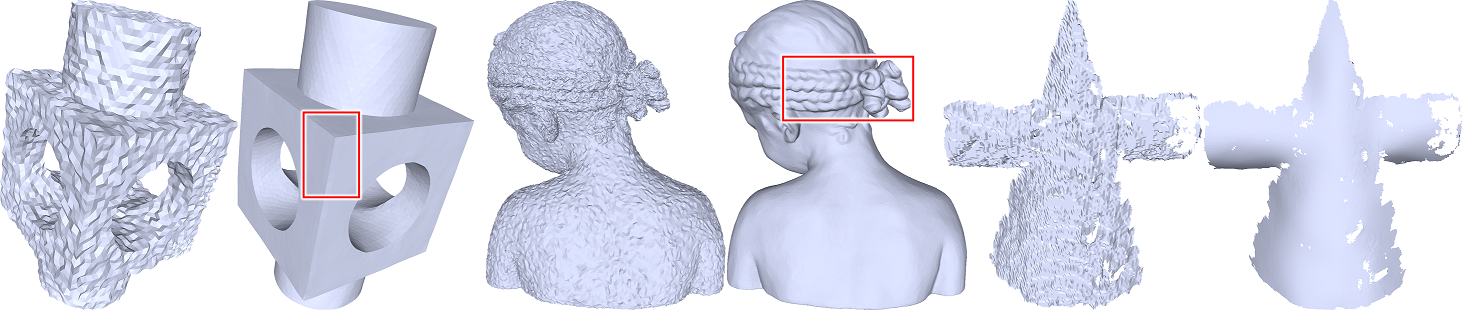}
	        \captionsetup{margin=0.8cm}
		\captionof{figure}{Our method is able to denoise meshes of various shapes and noise patterns, while preserving the fine details in the models; see the boxed regions in the above results.
		The left two models are corrupted by Gaussian noise, while the rightmost one is produced by Kinect v1 scans.
		Note that the input meshes are provided by~\cite{wang2016mesh}.}
		\label{fig:teaser}
	\end{center}
}

\IEEEtitleabstractindextext{%

\begin{justify}

\begin{abstract}
This paper presents a deep normal filtering network, called DNF-Net, for mesh denoising.
To better capture local geometry, our network processes the mesh in terms of local patches extracted from the mesh.
Overall, DNF-Net is an end-to-end network that takes patches of facet normals as inputs and directly outputs the corresponding denoised facet normals of the patches.
In this way, we can reconstruct the geometry from the denoised normals with feature preservation.
Besides the overall network architecture, our contributions include a novel multi-scale feature embedding unit, a residual learning strategy to remove noise, and a deeply-supervised joint loss function.
Compared with the recent data-driven works on mesh denoising, DNF-Net does not require manual input to extract features and better utilizes the training data to enhance its denoising performance.
Finally, we present comprehensive experiments to evaluate our method and demonstrate its superiority over the state of the art on both synthetic and real-scanned meshes.
\end{abstract}

\end{justify}

\begin{IEEEkeywords}
		Mesh denoising, normal filtering, deep neural network, data-driven learning, local patches.
\end{IEEEkeywords}}

\maketitle

\IEEEdisplaynontitleabstractindextext

% \fi
\IEEEpeerreviewmaketitle

%%%%%%%%% BODY TEXT
%%%%%%%%%%%%%%%%%%%%%%%%%%%%%%%%%%%%%%%%%%%%%%%%%%%%%%%%%%%%%%%%%%%%%%%%%%%%%%%%%%%%%%%
\section{Introduction}
\label{sec::introduction}

3D meshes are very common 3D representations widely-used in animations and games, as well as in various applications such as virtual and augmented reality, 3D simulations, medical shape analysis, etc.
While 3D meshes can be manually created by artists using software tools, the creation process is usually long and tedious.
Automatically capturing and reconstructing 3D meshes using scanning has become a viable and efficient solution for preparing 3D meshes.
However, raw meshes inevitably contain noise, so mesh denoising is often employed as a post-processing step to remove noise while preserving the fine object details.

Fundamentally, the key difficulty of mesh denoising lies on how to differentiate noise and fine details, which are both high frequency and small in scale~\cite{zheng2011bilateral,li2018non}.
In the literature, lots of efforts have been devoted to denoise meshes.
Traditional methods address the problem by introducing various kinds of filter-based models,~\ie, bilateral normal filtering~\cite{LeeWang05,zheng2011bilateral,zhang2015guided}, tensor voting~\cite{zhu2013coarse,wei2016tensor,yadav2017mesh}, and non-local low-rank normal filtering~\cite{li2018non,wei2018mesh}, or by assuming some kinds of priors,~\ie, $L_0$ minimization~\cite{LeiHe13}, $L_1$-norm sparsity~\cite{lu2017robust}, and $L_0$ sparse regularization~\cite{zhao2018robust}.
However, a noisy mesh may contain a variety of irregular structures that are corrupted by noise of different patterns.
Hence, making use of a particular filter or prior assumption to denoise meshes may not always produce satisfactory results.
Also, users often have to carefully fine-tune various model parameters in the methods for denoising different input meshes.

To circumvent these limitations, researchers began to explore data-driven methods~\cite{wang2016mesh,wei2018mesh}.
The basic idea of these methods is to regress functions that map noisy inputs to the ground-truth counterparts.
Although these pioneering methods are already data-driven, they still rely on manual inputs to extract features.
Hence, the valuable information available in the training data may not be fully exhausted.

\begin{figure*}[t]
	\centering
	\includegraphics[width=0.99\linewidth]{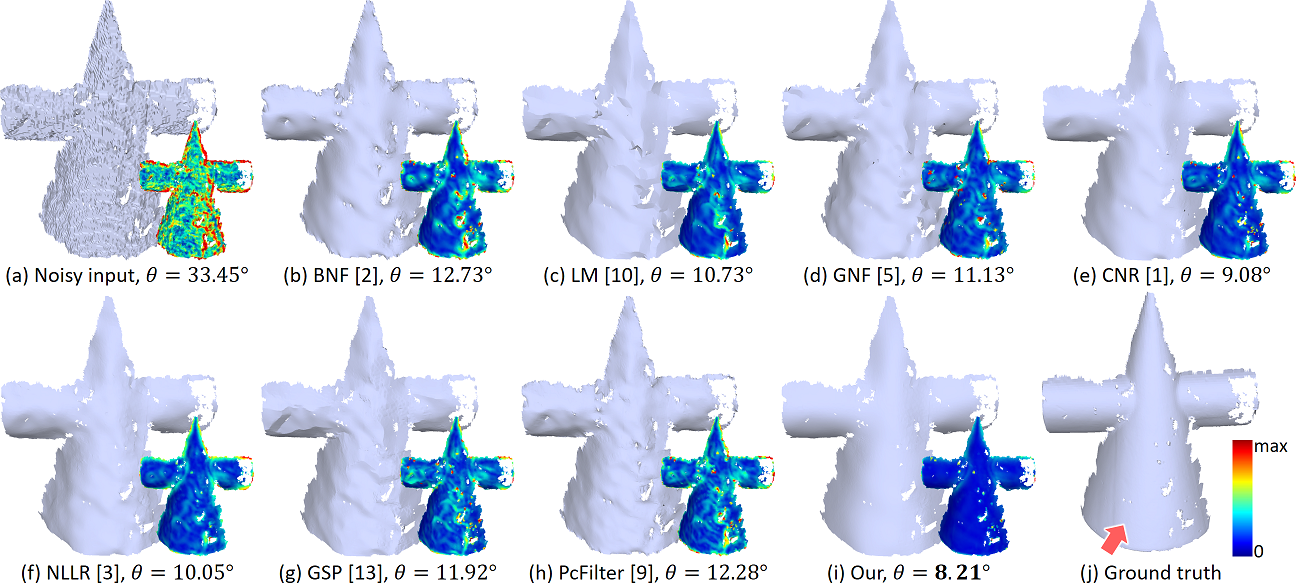}
	\caption{Comparing the performance of various methods (b)-(i) on denoising (a) an input noisy mesh scanned by Microsoft Kinect v1:
	BNF~\cite{zheng2011bilateral}({\footnotesize $\sigma_{s}$$=$$0.55$, $k_{\text{iter}}$$=$$50$, $v_{\text{iter}}$$=$$50$}),
	LM~\cite{LeiHe13} ({\footnotesize$\lambda$$=$$5$$\times$$10^{-5}$}),
	GNF~\cite{zhang2015guided} ({\footnotesize $\sigma_{r}$$=$$0.55$, $k_{\text{iter}}$$=$$30$, $v_{\text{iter}}$$=$$50$}),
	CNR~\cite{wang2016mesh},
	NLLR~\cite{li2018non} ({\footnotesize $\sigma_{\mathcal{M}}=0.65$, $v_{\text{iter}}=50$, $N_k=50$}),
	GSP~\cite{arvanitis2019feature},
	PcFilter~\cite{wei2018mesh},
	and our method, respectively.
	(j) The ground truth.
	We show also the associated normal error maps, where the colors reveal the angular difference between the corresponding normal vectors in ground truth and denoised meshes.
	Both the visual comparisons (compare the surface region marked by the red arrow above) and the $\theta$ values (mean angular difference; see Section~\ref{sec:experiments})
	show the superiority of our method over the others.}
	\label{fig:teaser2}
\end{figure*}

Unlike existing methods, we introduce a novel deep normal filtering network, called {\em DNF-Net\/}, for mesh denoising.
Given a noisy mesh with corrupted facet normals, our DNF-Net is able to robustly generate a corresponding denoised facet normal field, which is then employed to reconstruct the denoised mesh, while preserving the fine details, such as the sharp edges and corners, in the input mesh.

The key contribution in our method is a {\em deep neural network framework\/} that learns to filter normal vectors on meshes without requiring explicit information about the underlying surface or the noise characteristics.
To learn the local geometry patterns, our network processes the mesh in the form of patches on the mesh surface.
Particularly, to facilitate the network learning, we design the {\em multi-scale feature embedding unit\/} to extract the normal feature map, and the {\em residual learning unit\/} to regress the features of the noise per patch.
Also, we drive DNF-Net to learn by formulating the {\em deeply-supervised joint loss function\/}, which consists of a normal recovery loss and a residual regularization loss.

We performed several experiments to qualitatively and quantitatively evaluate DNF-Net.
Results show that DNF-Net is able to handle meshes of various shapes and noise patterns and produce high-quality denoised results for both synthetic and real-scanned noisy inputs in terms of denoising quality and feature preservation; see Figure~\ref{fig:teaser} for results produced by our method on various noisy meshes.
Figure~\ref{fig:teaser2} further shows a comparison example on a real-scanned noisy mesh, demonstrating the strong capability of DNF-Net to remove such severe noise, as compared with the various state-of-the-art methods; please see Section~\ref{sec:experiments} for more experiments and comparison results.

%%%%%%%%%%%%%%%%%%%%%%%%%%%%%%%%%%%%%%%%%%%%%%%%%%%%%%%%%%%%%%%%%%%%%%%%%%%%%%%%%%%%%%%
\section{Related Work}
\label{sec:related_work}

%%%%%%%%%%%%%%%%%%%%%%%%%%%%%%%%%%%%%%%%%%%%%%%%%%%%%%%%%%%%%%%

\begin{figure*}[t]
	\centering
	\includegraphics[width=0.99\linewidth]{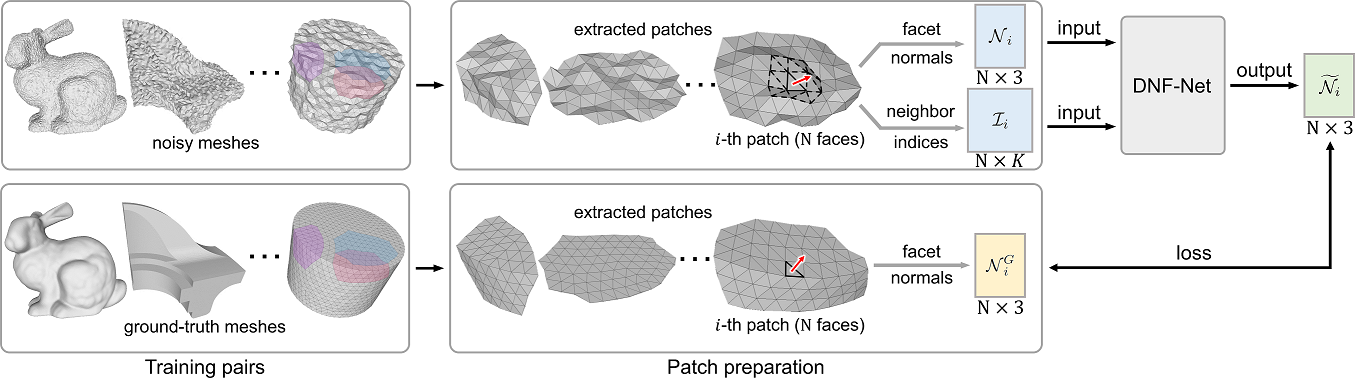}
	\caption{Illustration of our network training pipeline.
	Given a noisy mesh and its corresponding ground truth, we first crop local patches with $N$ faces.
	Then, for each noisy patch, we prepare a matrix of facet normals $\mathcal{N}_{i}$ and a matrix of local neighbor indices $\mathcal{I}_i$ as the network inputs.
	Also, we extract the corresponding ground-truth facet normals $\mathcal{N}^G_{i}$ and train the network to learn to directly output the denoised facet normals $\mathcal{\widetilde{N}}_i$, which is supervised by $\mathcal{N}^G_{i}$.}
	\label{fig:overview_method}
\end{figure*}

%%%%%%%%%%%%%%%%%%%%%%%%%%%%%%%%%%%%%%%%%%%%%%%%%%%%%%%%%%%%%%%

\subsection{Traditional methods for mesh denoising}
Early methods~\cite{field1988laplacian,taubin1995signal,desbrun1999implicit,vollmer1999improved} denoise meshes by formulating local isotropic filters to remove noise and solving the volume shrinkage problem caused by denoising.
Since the filter weights remain unchanged for varying surface characteristics, the denoised results are often overly smoothed.
Hence, various anisotropic techniques were proposed:

Fleishman~\etal~\cite{fleishman2003bilateral} and Jones~\etal~\cite{Jones03} extended the bilateral filtering technique in image denoising to mesh denoising to directly filter the vertex positions.
Later, observing that normal information can better capture the underlying surface characteristics, techniques based on bilateral normal filtering~\cite{LeeWang05,zheng2011bilateral,zhang2015guided} were introduced to {\em first filter facet normals and then adjust vertex coordinates accordingly\/}.
Very recently, Li~\etal~\cite{li2018non} and Wei~\etal~\cite{wei2018mesh} independently developed a non-local low-rank scheme to filter the normal field, where promising results were obtained.

Besides bilateral filtering and normal filtering, some works explore the notion of voting on the surface tensors to guide the mesh denoising process with feature preservation~\cite{zhu2013coarse,wei2016tensor,yadav2017mesh}.
Some other works formulate the mesh denoising problem as a global optimization and recover the meshes that best fit both the inputs and some pre-defined constraints or priors,~\eg, He~\etal~\cite{LeiHe13} explored $L_0$ minimization; Lu~\etal~\cite{lu2017robust} explored $L_1$-norm sparsity; and Zhao~\etal~\cite{zhao2018robust} explored $L_0$ sparse regularization.
However, these methods rely on priors of the noise distribution.
Recently, Arvanitis~\etal~\cite{arvanitis2019feature} proposed a novel coarse-to-fine graph spectral processing approach for mesh denoising.
Although the above methods work well for different noisy inputs, users have to specifically fine-tune various parameters to obtain satisfactory results for meshes of different geometry features and noise levels.

%%%%%%%%%%%%%%%%%%%%%%%%%%%%%%%%%%%%%%%%%%%%%%%%%%%%%%%%%%%%%%%

\subsection{Data-driven methods for mesh denoising}

There have been increasing attention on exploiting data-driven methods to denoise meshes.
Wang~\etal~\cite{wang2016mesh} presented a pioneering work called cascaded normal regression (CNR) to learn the mapping from noisy inputs to the ground-truth counterparts.
Considering that the learning process may lose some fine details, Wang~\etal~\cite{wang2019data} further developed a two-step data-driven method with the second step to enhance the recovery of the geometric details.

Although these data-driven approaches are able to learn the denoising pattern to a certain extent without specific assumptions on the underlying geometry features and noise patterns, they still need manually-extracted geometry descriptors from the noisy inputs,~\eg, the filtered facet normal descriptor~\cite{wang2016mesh}, without fully exploiting deep neural networks to automatically learn and extract features.
Hence, the information provided in the training data may not be fully exhausted.
To the best of our knowledge, this paper presents the first work that formulates a deep neural network to denoise meshes by filtering the raw facet normals.

%%%%%%%%%%%%%%%%%%%%%%%%%%%%%%%%%%%%%%%%%%%%%%%%%%%%%%%%%%%%%%%

\subsection{Deep neural networks for 3D model processing}
Driven by the success of deep learning in diverse computer vision, graphics, and image processing tasks, researchers in 3D geometry processing have started to explore deep neural networks for 3D model processing.
However, unlike 2D images with regular pixel grid structures, 3D models suffer from the property of irregular connectivity.
Hence, early works explored the transformation of the input 3D models to grid structures,~\eg,
volume representation~\cite{wu20153d,maturana2015voxnet,qi2016volumetric,wang2017cnn,wang2018adaptive},
depth map~\cite{qi2018frustum},
multi-view images~\cite{su2015multi,qi2016volumetric},~\etc, so that we can apply deep convolutional neural networks (CNNs) to directly process the data.

On the other hand, there have been extensive studies on directly taking deep neural networks to process 3D (irregular) point clouds.
PointNet~\cite{qi2017pointnet} and PointNet++~\cite{qi2017pointnet++} are two pioneering works that consume point clouds directly as the network input.
Subsequently, more network architectures were designed for handling various tasks on 3D point clouds, including
object recognition~\cite{li2018pointcnn,atzmon2018point,wang2018dynamic},
unsupervised feature learning~\cite{yang2018foldingnet,deng2018ppf},
upsampling~\cite{yu2018pu,yifan2018patch}, completion~\cite{yuan2018pcn}, instance segmentation~\cite{yi2019gspn},~\etc

Compared to point clouds, 3D meshes contain vertex connectivity in addition to point/vertex coordinates.
Hence, only a few works process 3D meshes using neural networks.
Some of them focus on generating meshes from single images~\cite{kato2018neural,wang2018pixel2mesh} or from incomplete range scans~\cite{litany2018deformable,dai2018scan2mesh}. Ranjan~\etal~\cite{ranjan2018generating} introduced a mesh autoencoder to generate 3D human faces.
Several studies on 3D shape representation directly operate on mesh data.
Hanocka~\etal~\cite{hanocka2019meshcnn} designed MeshCNN, a neural network that performs task-driven mesh simplification based on the edges between the mesh vertices.
Feng~\etal~\cite{feng2018meshnet} proposed MeshNet to learn from polygon faces for 3D shape classification and retrieval.
Yi~\etal~\cite{yi2017syncspeccnn} and Kostrikov~\etal~\cite{kostrikov2018surface} exploited the differential geometry properties of manifolds through Graph Neural Networks and its spectral extensions.
Different from these works, we design our DNF-Net to directly process normal vectors in local patches extracted on the mesh surface.
We do not assume a grid structure nor resample the normals into a grid; our network directly processes the normal vectors, as well as the local triangle connectivity information, on each patch as inputs and outputs the denoised normal vectors.

%%%%%%%%%%%%%%%%%%%%%%%%%%%%%%%%%%%%%%%%%%%%%%%%%%%%%%%%%%%%%%%%%%%%%%%%%%%%%%%%%%%%%%%
%\input{overview}
%%%%%%%%%%%%%%%%%%%%%%%%%%%%%%%%%%%%%%%%%%%%%%%%%%%%%%%%%%%%%%%%%%%%%%%%%%%%%%%%%%%%%%%
\section{Method}
\label{sec:method}

%%%%%%%%%%%%%%%%%%%%%%%%%%%%%%%%%%%%%%%%%%%%%%%%%%%%%%%%%%%%%%%%%%%%%%%%%
\begin{figure*}[t]
	\centering
	\includegraphics[width=0.99\linewidth]{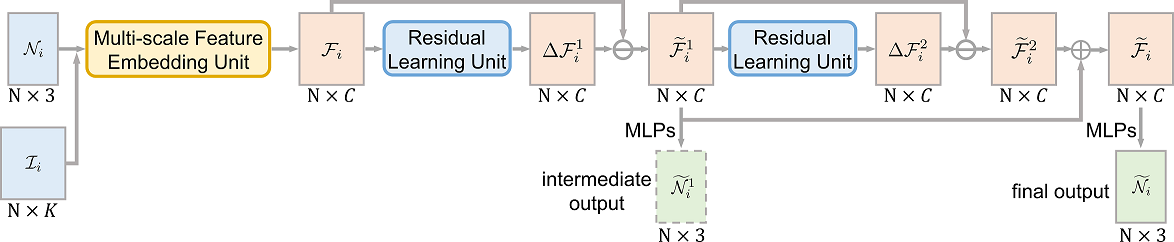}
	\caption{The overall network architecture of DNF-Net. We feed the patch facet normals $\mathcal{N}_{i}$ and corresponding neighbor indices $\mathcal{I}_i$ as network inputs to extract feature map $\mathcal{F}_i$ (with $C$ channels).
	Our DNF-Net removes noise by first learning the noise residual $\Delta \mathcal{F}_i$ from $\mathcal{F}_i$, then obtains the denoised feature map $\widetilde{\mathcal{F}}_i$ by subtracting the residual.
	The process is repeated to improve the noise removal.
	Lastly, we regress the output patch facet normals $\mathcal{\widetilde{N}}_i$ from the denoised feature map $\widetilde{\mathcal{F}}_i$.}
	\label{fig:framework}
	\vspace{-2mm}
\end{figure*}

%%%%%%%%%%%%%%%%%%%%%%%%%%%%%%%%%%%%%%%%%%%%%%%%%%%%%%%%%%%%%%%%%%%%%%%%%

\subsection{Overview}
\label{subsec:overview}

Given a noisy triangular 3D mesh $\mathcal{M}=(\mathbf{V}, \mathbf{F})$ with vertex set $\mathbf{V}$ and face set $\mathbf{F}$, our goal is to produce a denoised mesh $\widetilde{\mathcal{M}}=(\mathbf{\widetilde{V}}, \mathbf{F})$ from $\mathcal{M}$ with updated vertex set $\widetilde{\mathbf{V}}$.
Compared with vertex positions, first-order normal variations are known to better capture the local surface variations~\cite{tasdizen2003geometric}.
Therefore, we take a normal filtering approach~\cite{zheng2011bilateral,zhang2015guided,wang2016mesh,li2018non,wei2018mesh,wang2019data} to formulate our mesh denoising method.

Distinctively, we design a deep neural network, called {\em deep normal filtering network\/} (or {\em DNF-Net\/}) to learn to map the noisy facet normal vectors to noise-free ground-truth facet normal vectors on meshes.
In the course of formulating this network, we have the following considerations:
\begin{itemize}
\item
First, since mesh denoising is a low-level task, the network should focus on learning the local geometry.
Hence, we propose to crop {\em patches\/} on the object surface and process facet normals per patch in the network; see the left part in Figure~\ref{fig:overview_method}.
\item
Second, to enhance the generality of the network, it should abstract {\em local spatial patterns\/} instead of just encoding each facet normal individually.
Also, the network should produce the same results {\em regardless of the order in the normals in its input\/}; see $\mathcal{N}_{i}$ in Figure~\ref{fig:overview_method}.
Hence, based on the mesh connectivity, we extract {\em indices of the $K$-nearest faces\/} per face in the patch as the network inputs for the network to locate the face neighbors for further feature embedding.
\item
Lastly, for efficiency concern, our network is {\em end-to-end\/} to directly output denoised facet normals, and supervise the network training with corresponding facet normals from the ground-truth meshes.
\end{itemize}

In the following subsections, we first elaborate on the patch preparation procedure (Section~\ref{subsec:patch_prepare}).
We then introduce the architecture of DNF-Net and the loss function in the network training (Sections~\ref{subsec:architecture} \& \ref{subsec:loss}).
Lastly, we give details on the method implementation (Section~\ref{subsec:implementation}).

%%%%%%%%%%%%%%%%%%%%%%%%%%%%%%%%%%%%%%%%%%%%%%%%%%%%%%%%%%%%%%%%%%%%%%%%%

\subsection{Training patch preparation}
\label{subsec:patch_prepare}

Given a pair of meshes, a noisy mesh $\mathcal{M}$ and corresponding ground truth $\mathcal{M}^G$, as inputs, there are three steps to prepare the training patches from them.
First, we locate a set of faces on $\mathcal{M}$ as seeds to generate patches.
To randomly select seed faces, such that the resulting patches exhibit more diverse surface patterns, we calculate the one-ring facet normal variation around each face and randomly pick $P$ seed faces by an anisotropic sampling based on the normal variance.

Second, from each seed face, we grow a patch by finding the $N$$-$$1$ nearby faces on $\mathcal{M}$ with the shortest geodesic distances from the seed.
Specifically, to compute the geodesic distance between the seed face and a nearby face, we try each of its three vertices as the start point, find the shortest geodesic distance to each vertex of a nearby face, then take the smallest distance among the nine distances as the geodesic distance from the seed face to that nearby face.
Here, we use the heat method in~\cite{crane2013geodesics}.
Lastly, we sort all the distances among the surrounding (nearby) faces, and select the $N$$-$$1$ nearest ones.
Hence, we can produce patches (with $N$ faces) that are more regular in shape for training.

Lastly, we pack the $N$ facet normals on each patch as the patch normal matrix $\mathcal{N}_{i} \in \mathbb{R}^{N \times 3}$.
Also, we take advantage of the mesh connectivity and prepare patch index matrix $\mathcal{I}_i \in \mathbb{R}^{N \times K}$, where each row represents a face on the patch and stores the indices (row indices in $\mathcal{N}_{i}$) of the $K$-nearest faces to the face on the patch.
Then, we feed $\mathcal{N}_{i}$ and $\mathcal{I}_i$ as inputs to the network.
Further, for each patch formed on $\mathcal{M}$, we follow the same procedure to form a patch from the corresponding seed face on $\mathcal{M}^G$ and extract the corresponding $N$ facet normals to form matrix $\mathcal{N}^G_{i}$ as $\mathcal{N}_{i}$'s ground-truth to supervise the network; see Figure~\ref{fig:overview_method}.

%%%%%%%%%%%%%%%%%%%%%%%%%%%%%%%%%%%%%%%%%%%%%%%%%%%%%%%%%%%%%%%%%%%%%%%%%
\begin{figure*}[t]
	\centering
	\includegraphics[width=1.0\linewidth]{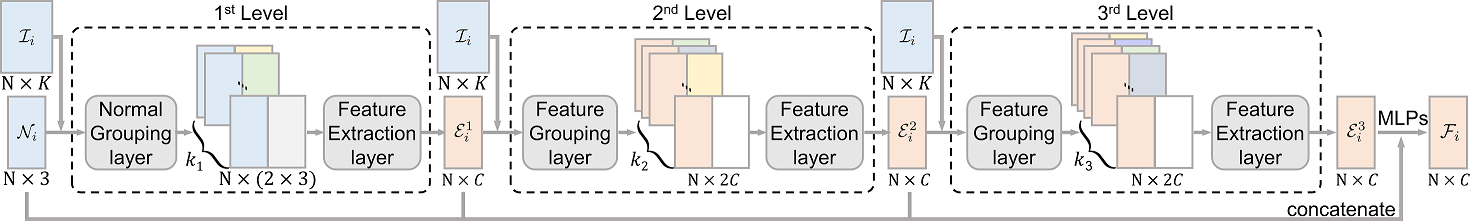}
	\caption{The three-level architecture of the multi-scale feature embedding unit.}
	\label{fig:embedding}
	\vspace*{-2.5mm}
\end{figure*}

%%%%%%%%%%%%%%%%%%%%%%%%%%%%%%%%%%%%%%%%%%%%%%%%%%%%%%%%%%%%%%%%%%%%%%%%%

\subsection{Network architecture}
\label{subsec:architecture}
Figure~\ref{fig:framework} shows the overall architecture of DNF-Net.
Taking $\mathcal{N}_i$
and $\mathcal{I}_i$
as inputs, DNF-Net first employs the multi-scale feature embedding unit (Section~\ref{subsub:embedding}) to extract the normal feature map $\mathcal{F}_i$.
Since $\mathcal{F}_i$ contains noise, we thus model it as $\mathcal{F}_i=\widetilde{\mathcal{F}}_i + \Delta \mathcal{F}_i$, where $\widetilde{\mathcal{F}}_i$ is the cleaned noise-free normal feature map and $\Delta \mathcal{F}_i$ is the feature map of the noise.

Considering that the underlying noise-free surface is usually more diverse compared with the noise patterns, thus encoding $\Delta \mathcal{F}_i$ is more effective than directly encoding $\widetilde{\mathcal{F}}_i$.
Hence, we feed $\mathcal{F}_i$ to a residual learning unit (Section~\ref{subsub:residual}) to first extract the residual $\Delta \mathcal{F}_i$ from $\mathcal{F}_i$.
Naturally, the cleaned feature map $\widetilde{\mathcal{F}}_i$ is recast into $\mathcal{F}_i - \Delta \mathcal{F}_i$.
To enhance the outputs, we cascade the residual-learning-and-subtraction process in a progressive manner to obtain an intermediate cleaned feature map $\widetilde{\mathcal{F}}^1_i$ in the middle of the network, besides the final cleaned feature map $\widetilde{\mathcal{F}}_i$; see Figure~\ref{fig:framework}.
Importantly, instead of only supervising the final output $\widetilde{\mathcal{N}}_i$ from $\widetilde{\mathcal{F}}_i$, we also regress an intermediate output $\widetilde{\mathcal{N}}^1_i$ from $\widetilde{\mathcal{F}}^1_i$ to give direct supervision when training the hidden layer in the network (Section~\ref{subsec:loss}).

%%%%%%%%%%%%%%%%%%%%%%%%%%%%%%%%%%%%%%%%%%%%%%%

\subsubsection{Multi-scale feature embedding unit}
\label{subsub:embedding}

For a comprehensive geometric understanding of a mesh structure, say locally around a vertex in the mesh, a general approach is to do a multi-scale analysis around the vertex, so that we can extract geometric features for different spatial scales.
Particularly, the geometric structures usually vary over scales.
Hence, given an input patch with $N$ facet normals, we formulate a multi-scale feature embedding unit of three levels to harvest geometric features of different scales. Specifically, the purpose of this unit is to extract normal feature map $\mathcal{F}_i \in \mathbb{R}^{N \times C}$ from $\mathcal{N}_i$ and $\mathcal{I}_i$, where $C$ is the number of channels and $N$ is the number of faces (normals) on input patch.
To do so, we build a three-level architecture to learn $\mathcal{F}_i$ by progressively enlarging the contextual scales; see Figure~\ref{fig:embedding} for the detailed illustration.

Specifically, in the first level of the multi-scale feature embedding unit, we design the \emph{normal grouping layer} and \emph{feature extraction layer} to generate an embedded feature map $\mathcal{E}_i^1 \in \mathbb{R}^{N \times C}$, given inputs $\mathcal{N}_i$ and $\mathcal{I}_i$.
In short, the normal grouping layer packs facet normals of the $k_1$ ($k_1$$<$$K$) nearest faces per face on the input patch using $\mathcal{I}_i$, while the feature extraction layer further extracts per-face local context information from the packed facet normals to learn $\mathcal{E}_i^1$.
In this way, the $C$-dimensional feature vector in each row of $\mathcal{E}_i^1$ encodes the local context around each face at a scale of $k_1$.
We shall elaborate on each layer later in this subsection.

The second level has a similar structure as the first level (see again Figure~\ref{fig:embedding}), but it replaces the normal grouping layer by the \emph{feature grouping layer}, since its input $\mathcal{E}_i^1$ is a feature map instead of a set of normals,~\ie, $\mathcal{N}_i$.
Also, it replaces $k_1$ by $k_2$ ($k_1$$<$$k_2$$<$$K$) to consider a larger local context to generate the next-level embedded feature map $\mathcal{E}_i^2 \in \mathbb{R}^{N \times C}$.
The third level is almost the same as the second level, but it considers an even larger local context with $k_3$ ($k_2$$<$$k_3$$\leq$$K$) to generate the embedded feature map $\mathcal{E}_i^3 \in \mathbb{R}^{N \times C}$ from $\mathcal{E}_i^2$.
Note that, to realize a progressively-enlarging local context to improve the feature embedding, we ensure $k_3 > k_2 > k_1$.
Lastly, we concatenate $\mathcal{N}_i$, $\mathcal{E}_i^1$, $\mathcal{E}_i^2$, and $\mathcal{E}_i^3$, and pass the result via a series of multi-layer perceptrons (MLPs) to generate $\mathcal{F}_i$; see the rightmost portion of Figure~\ref{fig:embedding}.

\begin{figure}[!t]
	\centering
	\includegraphics[width=0.99\linewidth]{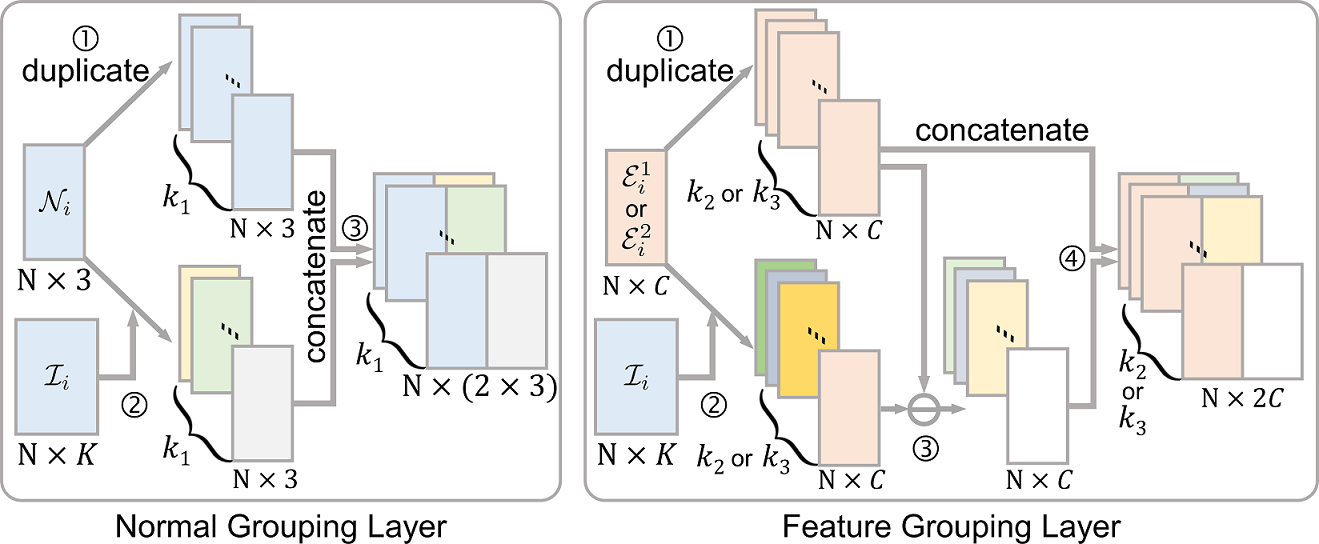}
	%\vspace{1mm}
	\caption{The normal grouping layer (left) and feature grouping layer (right) pack relevant local data (normals or features) for the feature extraction layer (see Figure~\ref{fig:embedding}) to process.}
	\label{fig:grouping}
	\vspace{-2mm}
\end{figure}

\para{Normal grouping layer}. \
To abstract local spatial patterns (features) around each face on the input patch, instead of just constructing per-face features by encoding individual facet normal, we design the normal grouping layer to pack nearby facet normal vectors around each face for subsequent feature extraction.
Note that each row in patch normal matrix $\mathcal{N}_i$ stores one normal vector of a face on the patch.

The detailed procedure of the normal grouping layer is illustrated on the left side of Figure~\ref{fig:grouping}.
Specifically, we first create $k_1$ duplicated copies of $\mathcal{N}_i$,~\ie, an $N \times k_1 \times 3$ volume.
Second, for each face in $\mathcal{N}_i$, we make use of $\mathcal{I}_i$ to locate the nearest $k_1$ faces to the face, then pack their facet normals to produce another $N \times k_1 \times 3$ volume, altogether for all the $N$ faces.
After that, we concatenate these two $N \times k_1 \times 3$ volumes to produce a new volume of size $N \times k_1 \times 6$, as the output of the normal grouping layer. %volume of data of size of
By doing so, we combine the global structure captured by the normal vectors over the patch, and the local neighborhood information captured by the packed nearby normals per face.
Overall, this normal grouping layer (similarly for the feature grouping layer) helps to pack, or re-arrange, the normal data, such that spatially-nearby normal vectors are grouped together per vertex.
By this means, we can consider local neighborhoods in the upcoming feature extraction layer and better capture local geometric structures.

\para{Feature grouping layer}. \
As shown in Figure~\ref{fig:embedding}, the inputs to the second and third levels are no longer 3D normal vectors but embedded feature vectors,~\ie, $\mathcal{E}_i^1$ or $\mathcal{E}_i^2 \in \mathbb{R}^{N \times C}$.
Hence, we design a feature grouping layer to pack these per-face embedded feature vectors; see the right side of Figure~\ref{fig:grouping}.

First, we also use $\mathcal{I}_i$ to locate the nearest $k_2$ (second level) or $k_3$ (third level) faces to generate an $N \times k_2 \times C$ or $N \times k_3 \times C$ feature volume.
However, rather than directly concatenating it with another volume of duplicated $\mathcal{E}_i^1$ or $\mathcal{E}_i^2$, we compute the residual via a subtraction operator before the feature volume concatenation; see step 3 on the right side of Figure~\ref{fig:grouping}.
Such subtraction helps to capture the local neighborhood information, while considering the global structure~\cite{wang2018dynamic}.
Note also that we employ the subtraction in feature grouping rather than normal grouping, since it helps to find residuals for embedded feature vectors but not for 3D normal vectors.
Hence, the output of the feature grouping layer is a volume of data of size of $N \times k_2 \times 2C$ for the second level or $N \times k_3 \times 2C$ for the third level.

\para{Feature extraction layer}. \
Feature extraction is important, since weak features offer less help to abstract the spatial structures, thus lowering the network performance.
See again Figure~\ref{fig:embedding}.
After normal grouping or feature grouping, we employ a feature extraction layer to extract features ($N$$\times$$C$) from the grouped normal vectors ($N$$\times$$k_1$$\times$$6$) or grouped feature vectors ($N$$ \times k_2 $$\times 2C$ or $N$$ \times k_3 $$\times 2C$).
A common solution here is to use MLPs followed by a max pooling, like several other works,~\eg,~\cite{qi2017pointnet++,yuan2018pcn,yu2018pu}.
However, since we employ the concatenation operation on the channel direction to combine both local and global information in our grouping layers (see Figure~\ref{fig:grouping}), we propose to use the channel attention module~\cite{woo2018cbam} to replace MLPs to better fuse the features among the different channels.
The basic idea of this module is to learn the channel weights from the grouped normal or feature vectors via MLPs, then use the weights to adjust the importance of each channel.
For more details of the channel attention module, readers may refer to~\cite{woo2018cbam}.

%%%%%%%%%%%%%%%%%%%%%%%%%%%%%%%%%%%%%%%%%%%%%%%

\subsubsection{Residual learning unit}
\label{subsub:residual}

After presenting the multi-scale feature embedding unit, we now go back to the overall architecture (see Figure~\ref{fig:framework}) and present the residual learning unit.
This unit extracts the noise feature $\Delta \mathcal{F}_i$ from the normal feature $\mathcal{F}_i$, so that we can later obtain the denoised feature map $\widetilde{\mathcal{F}}_i$ by $\mathcal{F}_i-\Delta \mathcal{F}_i$.

\begin{figure}[!t]
	\centering
	\includegraphics[width=0.85\linewidth]{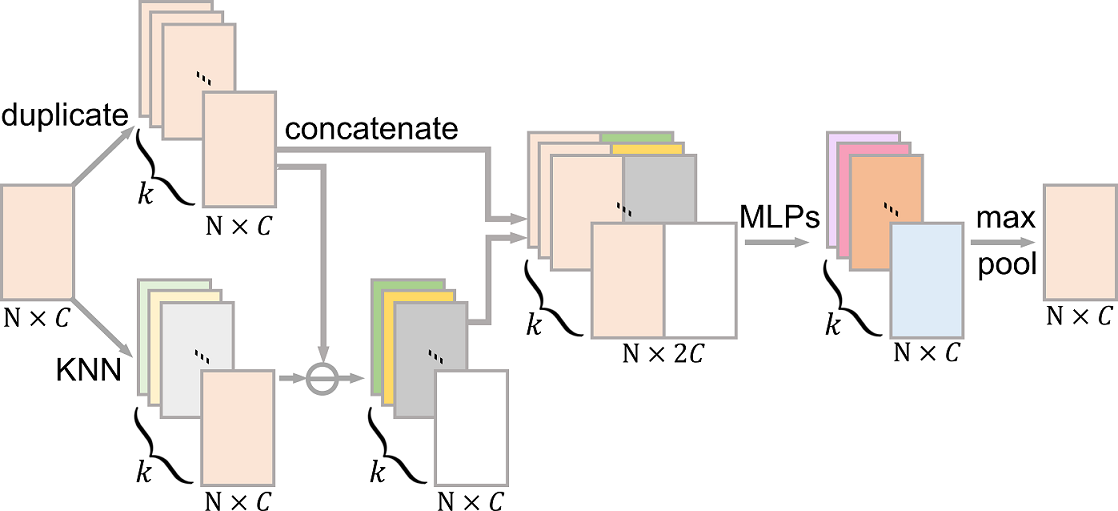}
	\caption{The architecture of the residual learning unit.}
	\label{fig:edge_conv}
	\vspace{-2mm}
\end{figure}

To better extract features for denoising, we should encode features over a local neighborhood rather than just as an individual feature vector.
Hence, like the feature grouping layer, this unit also finds the $k$-most similar feature vectors for each of the $N$ feature vectors in $\mathcal{F}_i$.
However, it employs KNN to locate similar feature vectors in the feature space instead of using the index matrix $\mathcal{I}_i$~\cite{wang2018dynamic}; see Figure~\ref{fig:edge_conv}.
The reason behind is that, in the multi-scale feature embedding unit, our goal is to extract representative features to encode local context, so using $\mathcal{I}_i$ enables us to locate features that are geodesically nearby.
In this residual learning unit, we, however, need to extract residual features from the input feature map $\mathcal{F}_i$, so we employ KNN search and extract features by considering feature similarity in the feature space.
In our experimental settings, as suggested by~\cite{wang2018dynamic}, we set $k=20$ in this unit.
As shown in Figure~\ref{fig:edge_conv}, after the concatenation between the duplicated ($k$ copies) input feature vectors ($\mathcal{F}_i$) and their associated $k$-most similar feature vectors, we use two MLPs followed by a max pooling to get the residual feature map $\Delta \mathcal{F}_i$ of size $N \times C$.

Also, as shown in Figure~\ref{fig:framework}, we use two consecutive residual learning units to progressively remove noise and to improve the overall denoising performance.
For an experiment that explores the effect of using a different number of residual learning units, please refer to Section~\ref{subsec:ablation}.

%%%%%%%%%%%%%%%%%%%%%%%%%%%%%%%%%%%%%%%%%%%%%%%%%%%%%%%%%%%%%%%%%%%%%%%%%

\subsection{Deeply-supervised end-to-end training}
\label{subsec:loss}

We design a deeply-supervised joint loss function with two terms to train the proposed network in an end-to-end manner:
(i) \emph{deeply-supervised normal recovery loss} and (ii) \emph{residual regularization loss}.

\para{Deeply-supervised normal recovery loss.} \
To encourage the denoised facet normals $\widetilde{\mathcal{N}}_i$ to be consistent with the ground truth normals $\mathcal{N}^G_i$, we use an $L_2$ norm to minimize the difference between $\widetilde{\mathcal{N}}_i$ and $\mathcal{N}^G_i$.
However, considering the deep-ness of our network, we further add another feedback, or supervision, on the companion intermediate output $\widetilde{\mathcal{N}}^1_i$ (see the middle part in Figure~\ref{fig:framework}), by applying an $L_2$ norm to minimize also the difference between $\widetilde{\mathcal{N}}^1_i$ and $\mathcal{N}^G_i$.
So, the deeply-supervised normal recovery loss is expressed as
\begin{equation}
\label{eq:deep}
	L_{\text{deep}} = \frac{1}{N_p} \sum_{i=1}^{N_p}(\|\mathcal{N}^G_i-\widetilde{\mathcal{N}}_i\|^2 \ + \|\mathcal{N}^G_i-\widetilde{\mathcal{N}}^1_i\|^2) \ ,
\end{equation}
where $N_p$ is the total number of training patches.
By doing so, we can directly influence the parameters in the hidden layer to enhance the quality of the feature maps.

\para{Residual regularization loss.} \
As shown in Figure~\ref{fig:framework}, DNF-Net progressively learns the residual features $\Delta \mathcal{F}^1_i$ and $\Delta \mathcal{F}^2_i$.
Theoretically, these residual features should just be a small portion of $\mathcal{F}_i$.
Having said that, the magnitude of $\Delta \mathcal{F}^1_i$ and $\Delta \mathcal{F}^2_i$ should not be too large.
Hence, we formulate the residual regularization loss on $\Delta \mathcal{F}^1_i$ and $\Delta \mathcal{F}^2_i$ as
\begin{equation}
\label{eq:reg}
	L_{\text{residual}} = \frac{1}{N_p} \sum_{i=1}^{N_p} (\|\Delta \mathcal{F}^1_i\|^2+\|\Delta \mathcal{F}^2_i\|^2) \ .
\end{equation}

\para{Joint loss.} \
Overall, we formulate the deeply-supervised joint loss function as a combination of Eqs.~\eqref{eq:deep} and~\eqref{eq:reg}:
\begin{equation}
L = L_{\text{deep}} + \alpha L_{\text{residual}} \ ,
\end{equation}
where $\alpha$ is a weight to balance the relative importance of the two loss terms, and we empirically set it as 0.5.

\subsection{Implementation details}
\label{subsec:implementation}

\para{Datasets}. \
In our experiments, we use the two benchmark datasets kindly provided by~\cite{wang2016mesh} to train our network: (i) a synthetic dataset and (ii) a real-scanned dataset.
The synthetic dataset contains 21 training models and 30 testing models, including CAD models, smooth models, and models with rich fine details.
For each model, the dataset provides a noise-free mesh as ground truth and three noisy meshes.
These noisy meshes are generated by adding three different magnitudes of Gaussian noise into the noise-free mesh; the standard deviations of the noise in these meshes are $0.1\bar{l}_e$, $0.2\bar{l}_e$, and $0.3\bar{l}_e$, where $\bar{l}_e$ is the average edge length in the mesh.
On the other hand, the real-scanned dataset has 146 training noisy meshes and 149 testing noisy meshes.
Each noisy mesh is accompanied with a clean mesh as the ground truth.
For more details, readers may refer to~\cite{wang2016mesh}.

\begin{figure*}[t]
	\centering
	\includegraphics[width=1.0\linewidth]{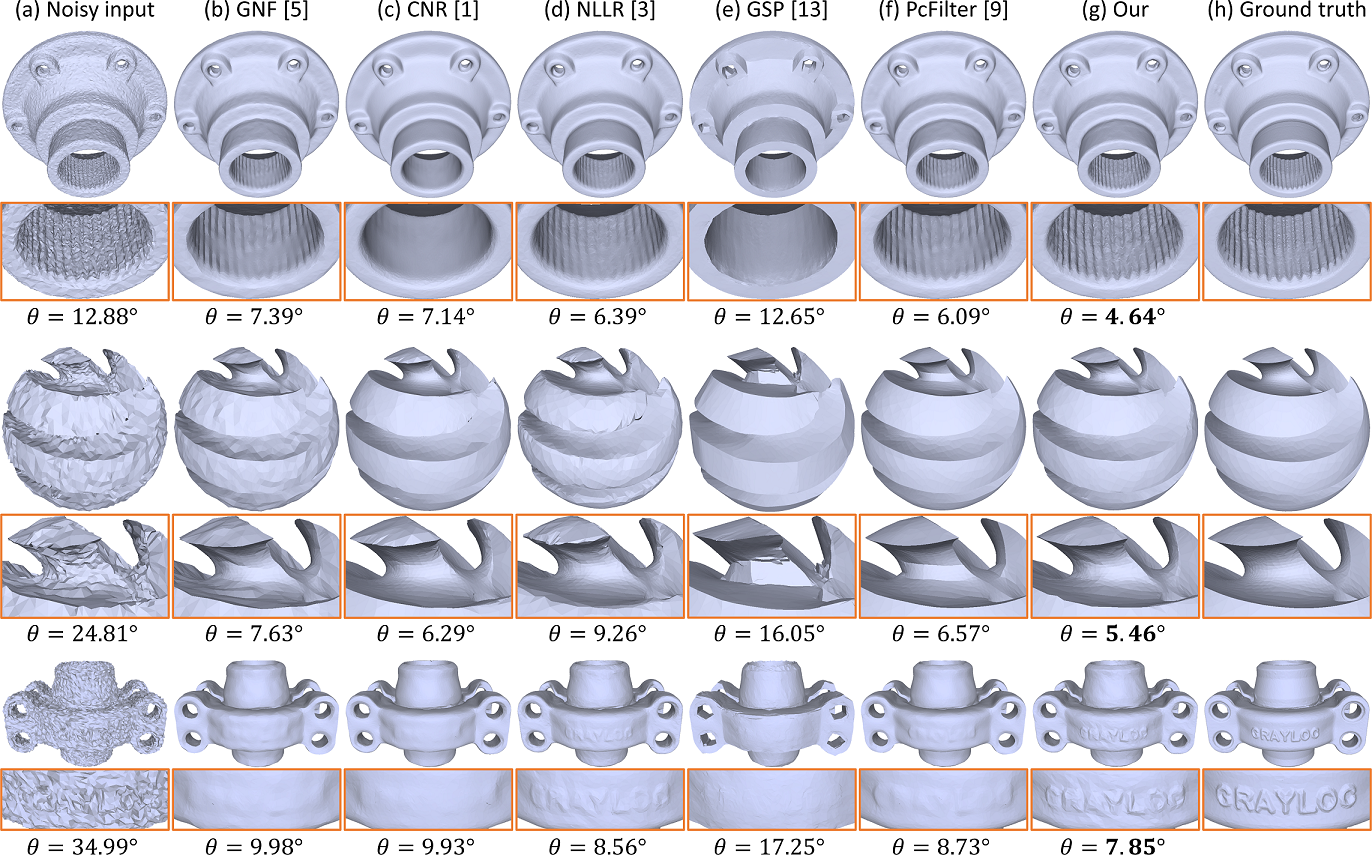}
	%\vspace{1mm}
	\caption{Comparing the mesh denoising results produced using different methods (b)-(g) on the synthetic noisy models shown on the leftmost column; these inputs have different Gaussian noise levels: $0.1\bar{l}_e$, $0.2\bar{l}_e$, and $0.3\bar{l}_e$ (top to bottom).}
	\label{fig:vis_synthetic}
	\vspace{-2mm}
\end{figure*}

\para{Network training}. \
Observing that the synthetic and real-scanned datasets have very different noise distributions, we follow the existing data-driven method, i.e., CNR~\cite{wang2016mesh}, to train our network separately on each dataset for obtaining our results on synthetic meshes and real-scanned meshes.
We plan to explore knowledge distillation techniques in the future to combine the two trained network models.

For each training mesh in the synthetic dataset, we crop $P$$=$$100$ patches, each with $N$$=$$800$ faces, such that a patch roughly covers 5\% to 10\% of the whole mesh.
For each face in a patch, we empirically store $K$$=$$50$ neighboring face indices, and uniformly set $k_1$$=$$10$, $k_2$$=$$30$, and $k_3$$=$$50$ in the multi-scale feature embedding unit (see Section~\ref{subsub:embedding}) of our network.
For each training mesh in the real-scanned dataset, we crop more patches with $P$$=$$200$, due to the complex noise distributions, and each patch has $N$$=$$800$ faces.
Considering that most of the real-scanned meshes have simple and smooth structures, we store $K$$=$$150$ neighbor facet indices for each face and set $k_1$$=$$50$, $k_2$$=$$100$, and $k_3$$=$$150$.
Please see the supplementary material for the effect of different parameter settings on the real-scanned dataset.

To avoid over-fitting in network training, we augmented $\mathcal{N}_{i}$ by adopting random rotation and jittering.
We implemented our method using TensorFlow and adopted the Adam optimizer with a mini-batch size of 10 and a learning rate of 0.001, as well as trained the network with 400 epochs.
Our trained network model, data, and code can be found on the GitHub project page~\footnote{https://github.com/nini-lxz/DNF-Net}.

\para{Network inference.} \
Given a test mesh, we crop patches on the mesh following the procedure of preparing training patches, then employ the trained network to produce denoised normals on the patches.
After that, we integrate the denoised normals on all patches to compute facet normals over the mesh, and follow~\cite{li2018non} to pass the restored facet normals to the iterative vertex updating method~\cite{sun2007fast} to update vertex positions and produce the denoised mesh.

%%%%%%%%%%%%%%%%%%%%%%%%%%%%%%%%%%%%%%%%%%%%%%%%%%%%%%%%%%%%%%%%%%%%%%%%%%%%%%%%%%%%%%%

\section{Results and Discussions}
\label{sec:experiments}

To demonstrate the effectiveness of our method, we compare it with several state-of-the-art methods, including
$L_0$ minimization (LM)~\cite{LeiHe13},
guided normal filtering (GNF)~\cite{zhang2015guided},
cascaded normal regression (CNR)~\cite{wang2016mesh},
non-local low-rank normal filtering (NLLR)~\cite{li2018non},
graph spectral processing approach (GSP)~\cite{arvanitis2019feature},
and patch normal co-filter (PcFilter)~\cite{wei2018mesh}.
For LM, GNF, and NLLR, we obtained their publicly-released codes and fine-tuned their model parameters with best effort to produce their denoising results; see our supplementary material for the details of the employed parameter values.
For CNR, we directly employed their released trained models to generate their results, while
for GSP and PcFilter, we obtained their results directly from the authors.

Following the recent works~\cite{wang2016mesh,li2018non}, we also employed the mean angular difference metric (denoted as $\theta$) to quantitatively evaluate and compare the results produced by the various methods.
By definition, $\theta$ is the mean angular difference (in degrees) between the corresponding facet normals in the ground truth and denoised meshes.
Hence, a small $\theta$ value indicates a better denoising result.
Note that, $\theta$ is calculated on each denoised mesh after the vertex update, and all methods being compared (including our method) use the same vertex update algorithm~\cite{sun2007fast}.

\begin{figure}[h]
	\centering
	\includegraphics[width=0.95\linewidth]{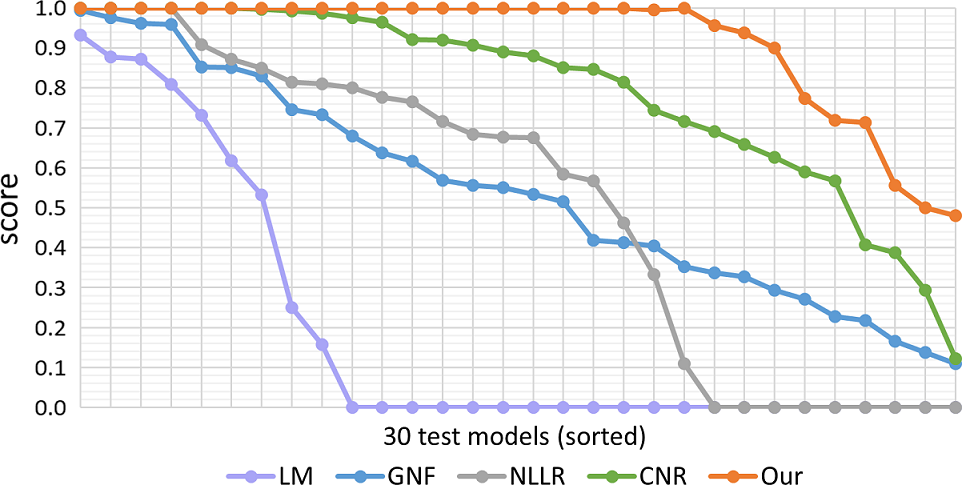}
	%\vspace{1mm}
	\caption{Comparing the scores obtained by different methods over 30 test models (sorted in descending order of score).}
	\label{fig:synthetic_plot}
	\vspace{-2mm}
\end{figure}

\begin{figure*}[t]
	\centering
	\includegraphics[width=0.99\linewidth]{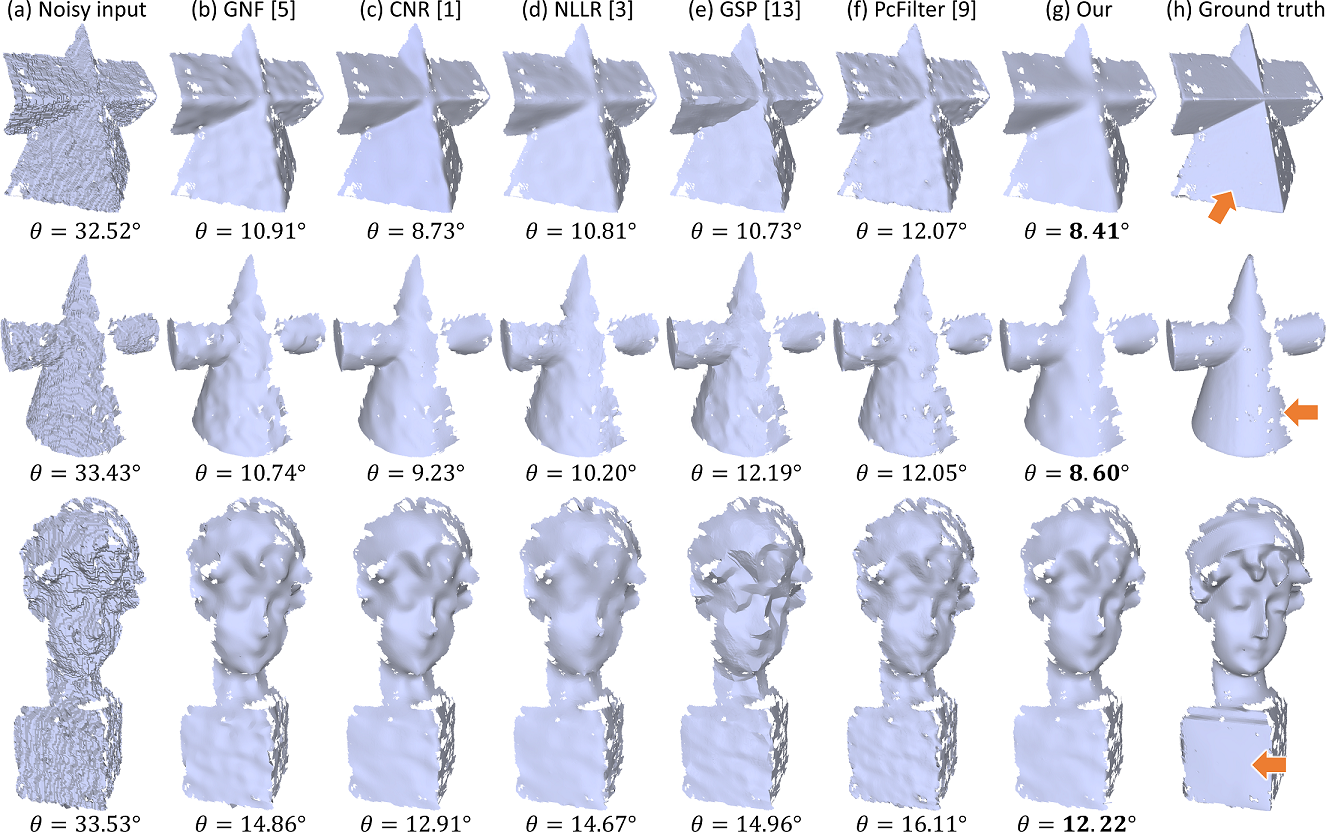}
	%\vspace{1mm}
	\caption{Comparing the mesh denoising results produced using different methods (b)-(g) on real-scanned noisy models.}
	\label{fig:vis_real}
\end{figure*}

%%%%%%%%%%%%%%%%%%%%%%%%%%%%%%%%%%%%%%%%%%%%%%%%%%%%%%%%%%%%%%%%%%%%%

\subsection{Results on Synthetic Models}
\label{subsec:synthetic}
First, we compare our method with the state-of-the-arts on test models provided in the synthetic dataset of~\cite{wang2016mesh}.
Figure~\ref{fig:vis_synthetic} shows the visual comparisons on three noisy meshes with different amount of Gaussian noise.
Comparing the results produced by our method (g) and others (b-f) with the ground truths (h), we can see that the other methods tend to over-smooth the fine details, over-sharpen the edges, or retain excessive noise in the results.
Our method is able to preserve more geometric details and recover the sharp edges, while effectively removing the noise; see particularly the blown-up views in Figure~\ref{fig:vis_synthetic}.
See also the $\theta$ value below each generated result.
Overall, our method achieved the smallest $\theta$ values for all models compared with all the other methods; see Part 1 of the supplementary material for more comparison results.
Besides, for each denoised mesh shown in Figure~\ref{fig:vis_synthetic}, we visualize its normal error distribution; please refer to Part 9 of our supplementary material.

%%%%%%%%%%%%%%%%%%%%%%%%%%%%%%%%%%%%%%%%%%%%%%%%%%%%%%%%%%%%%%%%%%%%%

\newcommand{\BE}[1]{{\textbf{#1}}}

Further, we expand the quantitative comparison by considering all the 30 test models in the dataset.
Note that, for each test model, \cite{wang2016mesh} provides three noisy meshes with different Gaussian noise levels.
Here, we define $\bar{\theta_j^i}$ as the averaged $\theta$ value achieved by method $i$ over the three noisy meshes of the $j$-th test model.
Also, we define normalized score $\text{score}_j^i \in [0,1]$ achieved by method $i$ on $j$-th test model as
\begin{equation}
\text{score}_j^i = 1 - \frac{\bar{\theta_j^i} - \min_i \bar{\theta_j^i}}{\max_i \bar{\theta_j^i} - \min_i \bar{\theta_j^i}} \ ,
\end{equation}
so a value of one indicates best result among the methods.

Figure~\ref{fig:synthetic_plot} plots the scores achieved by each method over the 30 test models.
Note that to more clearly reveal the results, the 30 score values over the test models per method are sorted in descending order when producing each plot.
Also, we consider only LM~\cite{LeiHe13}, GNF~\cite{zhang2015guided}, CNR~\cite{wang2016mesh}, and NLLR~\cite{li2018non}, since we have obtained their code or denoising results for all the test models in the dataset.
From the plots, we can see that our method obtains the best results (scores of one) for most test models, while the other methods start to drop much earlier.
More importantly, the test models have various geometric structures, including simple and smooth surface, sharp edges, and highly-detailed fine structures.
The reason behind the success of our method is that formulating a deep neural network allows us to more exhaustively extract discriminative features from the data to differentiate the underlying details and noise in the input meshes, while the other methods heavily rely on hand-crafted features from the assumptions or priors on the input models.

%%%%%%%%%%%%%%%%%%%%%%%%%%%%%%%%%%%%%%%%%%%%%%%%%%%%%%%%%%%%%%%%%%%%%

\subsection{Results on Real-scanned Models}
Next, we compare our method with others on real-scanned models.
Besides the results shown earlier in Figure~\ref{fig:teaser2}, we further show in Figure~\ref{fig:vis_real} more visual comparison results on three other Kinect real-scanned models~\cite{wang2016mesh}.
From the noisy input models on the leftmost column of the figure, we can see that Kinect scanning produces severe and irregular noise.
Such noise pattern differs from that of the Gaussian noise.
Comparing the results produced by various methods, the other methods tend to retain noise in their results, or fail to smooth flat surfaces.
On the contrary, our method is able to produce denoised models that are more smooth and closer to the ground truths, as verified again by the smallest $\theta$ values achieved by our method on the models.
Additionally, please refer to Parts 2 and 9 of the supplementary material for more real-scanned comparison results and for the normal error visualizations on the denoised meshes shown in Figure~\ref{fig:vis_real}, respectively.

\begin{table}[t]
	\caption{Comparing the denoising quality of our full pipeline with various cases in the ablation study.}
	\label{tab:ablation}
	\centering
	\vspace*{-2mm}
	\begin{center}
		\resizebox{1.0\linewidth}{!}{%
			\begin{tabular}{C{1.4cm}|C{1.15cm}C{1.15cm}C{1.15cm}C{1.15cm}C{1.15cm}C{1.15cm}}
				\toprule[1pt]
				Different & \multicolumn{5}{c}{Model} \\ \cline{2-6}
				cases & Block & Cube & Sphere & \hspace*{-2mm}Carter100K & Eros100K\\
				\cline{1-6}
				\emph{w/o inter}
				& 2.38$^{\circ}$
				& 1.38$^{\circ}$
				& 2.38$^{\circ}$
				& 5.84$^{\circ}$
				& 7.21$^{\circ}$
				\\
				\emph{w/o reg}
				& 2.43$^{\circ}$
				& 1.13$^{\circ}$
				& 2.43$^{\circ}$
				& 5.88$^{\circ}$
				& 7.20$^{\circ}$
				\\
				\emph{w/o sub}
				& \ \hspace*{-1.55mm} 2.69$^{\circ}$
				& \ \hspace*{-1.55mm} 1.42$^{\circ}$
				& \ \hspace*{-1.55mm} 2.41$^{\circ}$
				& \ \hspace*{-1.55mm} 6.02$^{\circ}$
				& \ \hspace*{-1.55mm} 7.23$^{\circ}$
				\\
				\emph{1 res unit}
				& 2.24$^{\circ}$
				& 1.25$^{\circ}$
				& 2.41$^{\circ}$
				& 5.92$^{\circ}$
				& 7.23$^{\circ}$
				\\
				\emph{3 res units}
				& 2.12$^{\circ}$
				& 1.03$^{\circ}$
				& \BE{2.24$^{\circ}$}
				& 5.86$^{\circ}$
				& \BE{7.14$^{\circ}$}
				\\ \cline{1-6}
				\emph{full pipeline}
				& \BE{2.12$^{\circ}$}
				& \BE{0.96$^{\circ}$}
				& 2.37$^{\circ}$
				& \BE{5.81$^{\circ}$}
				& 7.20$^{\circ}$
				\\
				\bottomrule[1pt]
		\end{tabular}}
	\end{center}
\end{table}

%%%%%%%%%%%%%%%%%%%%%%%%%%%%%%%%%%%%%%%%%%%%%%%%%%%%%%%%%%%%%%%%%%%%%

\subsection{Network Ablation Study}
\label{subsec:ablation}
To evaluate the effectiveness of the major components in our method, we conducted an ablation study by simplifying DNF-Net in the following four cases.
\begin{itemize}
\item[(i)]
\emph{w/o inter}: we remove the supervision on intermediate output $\widetilde{\mathcal{N}}^1_i$ from the first residual learning unit in the network (see Figure~\ref{fig:framework}), and supervise only the final output $\widetilde{\mathcal{N}}_i$,~\ie, we remove the second term in Eq.~\eqref{eq:deep}.
\item[(ii)]
\emph{w/o reg}: we remove the residual regularization loss term (\ie, Eq.~\eqref{eq:reg}) from the total loss of our network.
\item[(iii)]
\emph{w/o sub}: we compute a residual in step 3 of the feature grouping layer (Figure~\ref{fig:grouping}) by a subtraction operation before concatenating two feature volumes.
Such an operation helps capture the local information for denoising.
To verify its effectiveness, we remove it and directly concatenate the two feature volumes.
\item[(iv)]
\emph{$<$\#$>$ res units}:
as shown in Figure~\ref{fig:framework}, our network cascades two residual learning units successively in the overall network architecture.
Instead of deploying two residual learning units, we tried ``\emph{1 res unit}'' and ``\emph{3 res units}'' to explore the effect of the number of residual learning units on the network performance.
Note that, for fair comparison, we modified both Eqs.~\eqref{eq:deep} \&~\eqref{eq:reg} to supervise the output of every residual learning unit in the network.
\end{itemize}

\begin{figure}[!t]
	\centering
	\includegraphics[width=0.9\linewidth]{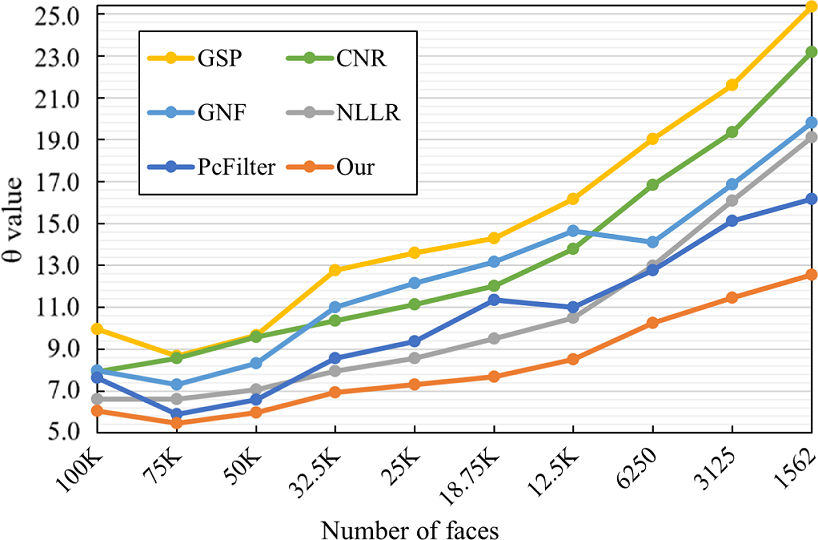}
	%\vspace{1mm}
	\caption{Denoising performance ($\theta$) of different methods on the same test model in different resolutions.
		Note that we start with the Eros model with 100K faces ($0.1\bar{l}_e$ Gaussian noise) and use the quadric edge collapse decimation method in MeshLab~\cite{journals/ercim/CignoniCR08} to progressively simplify the mesh model.}
	\label{fig:resolution_chart}
	\vspace{-2mm}
\end{figure}

\begin{figure*}[t]
	\centering
	\includegraphics[width=0.92\linewidth]{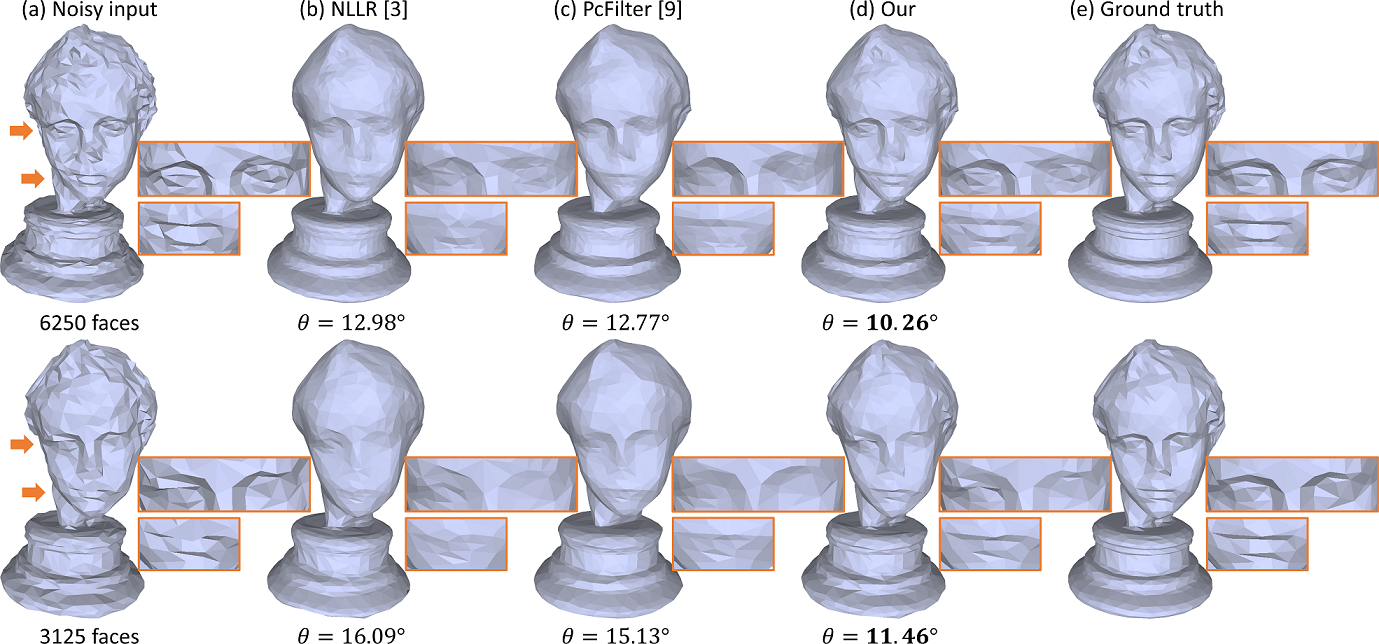}
	\vspace{-1mm}
	\caption{Comparing the denoising results produced by various methods (b)-(d) on two low-resolution noisy inputs (a) of the same model.
		Note that the input meshes were corrupted by Gaussian noise with a standard deviation of $0.1\bar{l}_e$.}
	\label{fig:resolution}
	\vspace{-2mm}
\end{figure*}

Specifically, we re-trained the network model separately for each case using the same training dataset of synthetic models (see Section~\ref{subsec:implementation}) and tested each network on five test models,~\ie, Block, Cube, Sphere, Carter100K, and Eros100K, which contain sharp edges, simple structure, and fine details.
As mentioned earlier, each provided model has three versions of noisy meshes.
Hence, similar to Section~\ref{subsec:synthetic}, we test each network on all the three noisy meshes per model, then compute the averaged $\theta$ value achieved over the three noisy meshes as the overall $\theta$ value of the model.

Table~\ref{tab:ablation} shows the results.
By comparing the top three rows with the bottom-most row (our full pipeline), we can see that each term in our loss function contributes to the mesh denoising performance, including the supervision on intermediate output and residual regularization loss term.
Besides, the subtraction operation in the feature grouping layer also contributes to improving the overall performance.

Further, the network with only one residual learning unit (4-th row) achieved a worse result than our full pipeline with two residual learning units.
If we increase the number of residual learning units to three (5-th row), the performance improves only slightly on two models, but the number of network parameters increases from 0.36M to 0.44M.
Hence, we decided to deploy two residual learning units in our network to balance the performance and efficiency.

\subsection{Robustness Test}

%In this section, we explore the stability and effectiveness of our DNF-Net under two extreme cases: (i) decreasing facet number; and (ii) non-uniform mesh triangulation.

Next, we explore the robustness of DNF-Net by considering (i) varying mesh resolution (\ie, different number of faces in the same model); (ii) irregular mesh triangulation; (iii) unseen noise patterns; and (iv) varying noise intensities.

\vspace*{1mm}
\para{Robustness on mesh resolution.} \
Most existing methods are sensitive to the mesh resolution, where the error metric $\theta$ could have large fluctuation for low-resolution meshes.
It is because these methods filter normals (or vertices) using a local neighborhood of a fixed number of rings,~\eg, one- or two-ring neighborhood.
Hence, when given a low-resolution mesh, they could involve a too large neighborhood in the filtering, thus leading to over-smoothing~\cite{zhang2015guided}.

Thanks to the patch-based training strategy, our DNF-Net is trained for various sizes of local neighborhoods for different models.
Hence, it can become less sensitive to the mesh resolution.
To explore this, we employed noisy Eros models of different resolutions as the inputs and plotted $\theta$ of the denoised results by various methods in Figure~\ref{fig:resolution_chart}.
Here, we directly used our trained network to test all the inputs without re-training or fine-tuning the network.
From the plots in Figure~\ref{fig:resolution_chart}, we can see that
our method (orange plot) consistently has the lowest $\theta$ values vs. all the others for various mesh resolutions.
Particularly, when the number of faces is reduced to less than 50K, the gaps of $\theta$ values between our method and the others gradually increase.
For example, when the number of faces is reduced from 50K down to 1562, the increased $\theta$ value for our method is 6.57 vs. 9.58 for the second-best one.
This shows that DNF-Net is not as sensitive as others to the mesh resolution.

Further, Figure~\ref{fig:resolution} shows the visual comparisons on two example low-resolution Eros models with 6250 and 3125 faces.
Comparing the results in (b), (c), and (d), we can see that the other two methods overly smooth out the details, while our method can better preserve the fine details; see the areas around the eyes and mouth in the results.
Also, it achieves the lowest $\theta$ values for both resolutions.

\begin{figure*}[t]
	\centering
	\includegraphics[width=0.99\linewidth]{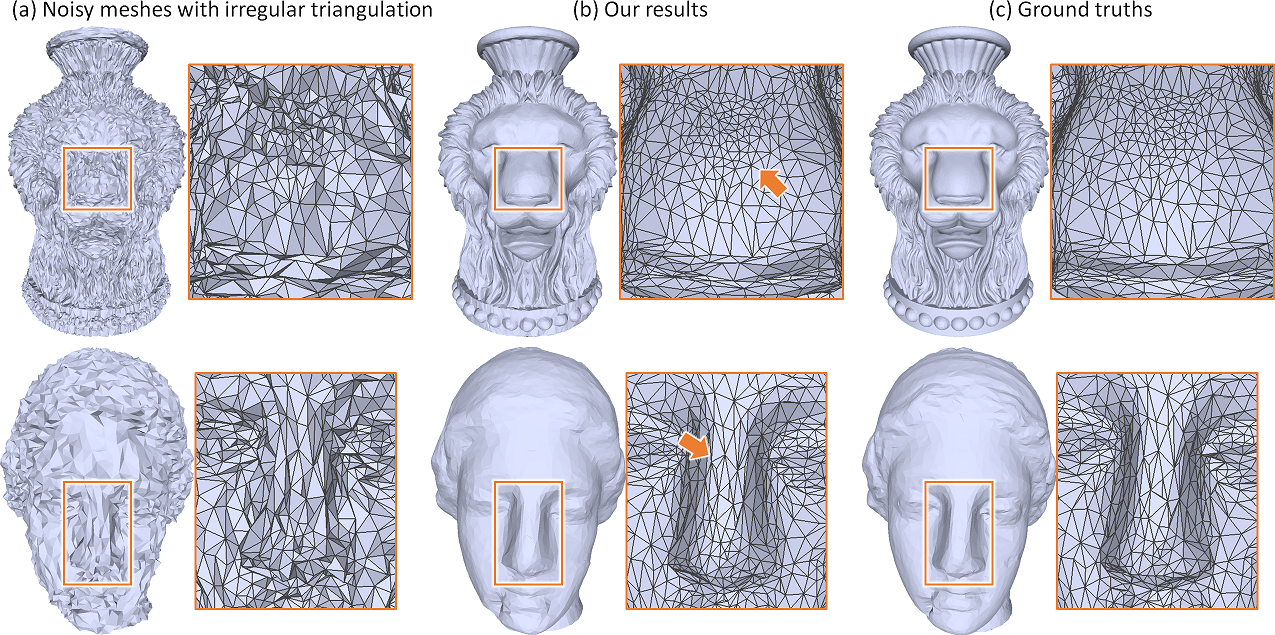}
	%\vspace{1mm}
	\caption{Denoising results produced by our DNF-Net on two irregular triangulated meshes with $0.3\bar{l}_e$ Gaussian noise.
			Apparently, the triangles around the nose regions are elongated with irregular vertex degrees, and our method can still produce results that are quite similar to the ground truths.}
	\label{fig:uniform}
\end{figure*}

\vspace*{1mm}
\para{Robustness on irregular triangulation.} \
Next, we test the robustness of our method on handling noisy meshes with irregular triangulation.
Figure~\ref{fig:uniform} (a) shows two example meshes with elongated triangles and irregular vertex degrees; see particularly the blown-up views of the nose regions in the figure.
Still, the results produced by our method (b) are quite close to the ground truths (c).
Note that some comparison results with other methods are also presented in Part 8 of the supplementary material.

\begin{figure}[t]
	\centering
	\includegraphics[width=0.8\linewidth]{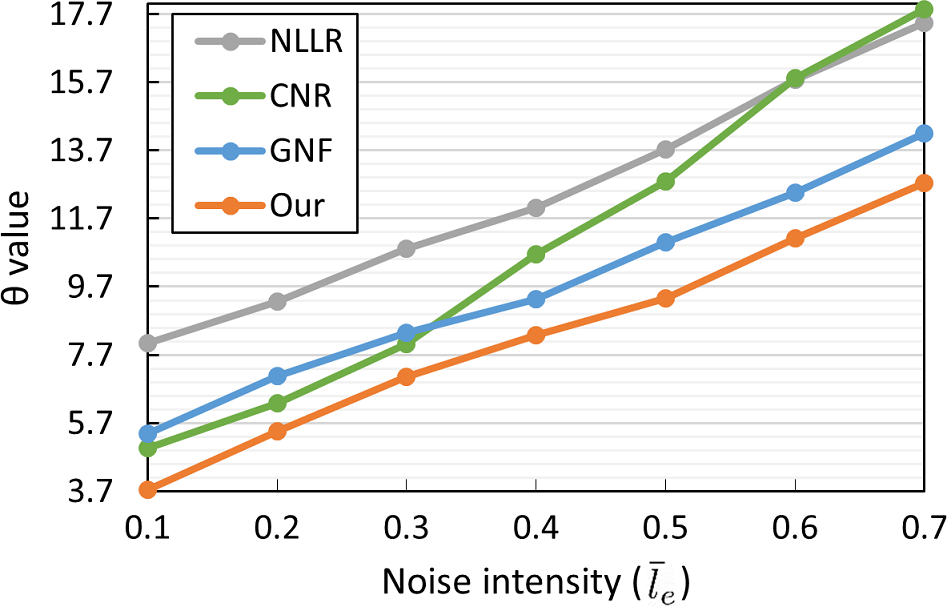}
	%\vspace{1mm}
	\caption{Comparing the mesh denoising performance of various methods on input meshes of increasing noise intensity.}
	\label{fig:noise_level}
	\vspace{-2mm}
\end{figure}

\begin{figure*}[t]
	\centering
	\includegraphics[width=0.99\linewidth]{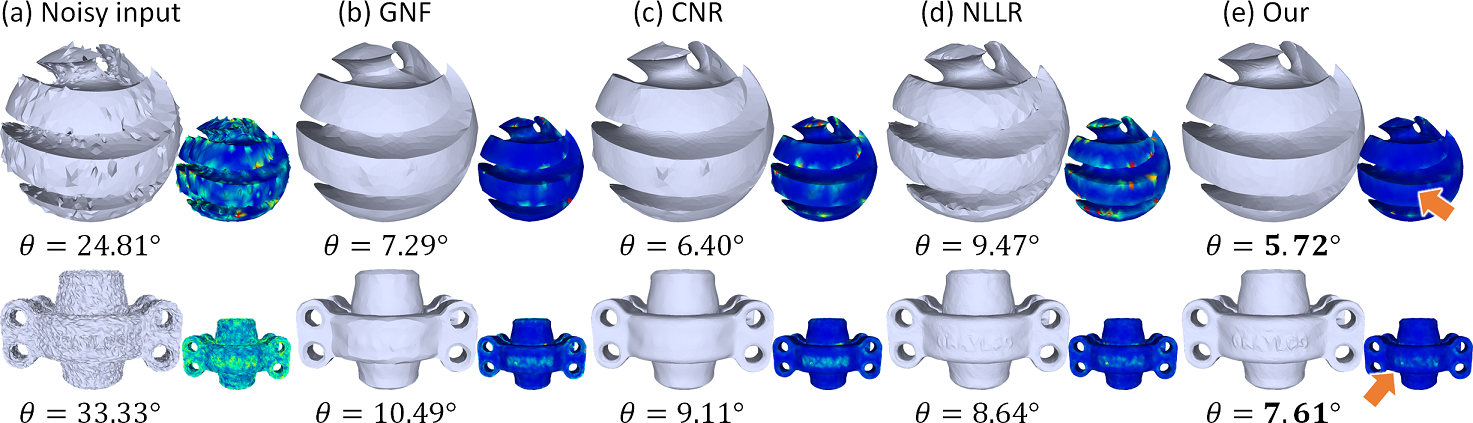}
	%\vspace{1mm}
	\caption{Comparing the mesh denoising performance of various methods (b)-(e) on two synthetic models that are corrupted by impulsive noise (top) and uniform noise (bottom).}
	\label{fig:noise_pattern}
	\vspace{-2mm}
\end{figure*}

\para{Robustness on noise patterns.} \
Further, we tested the generalization ability of our network on handling 3D meshes of noise patterns different from the noise pattern employed in the training set.
To do so, we directly employed our trained network to denoise meshes corrupted by impulsive noise and uniform noise (by replacing the Gaussian noise we employed with a uniform-distributed noise).
Figure~\ref{fig:noise_pattern} shows comparisons on two 3D meshes corrupted with impulsive noise (top) and uniform noise (bottom).
From the normal error visualizations and the $\theta$ values presented in the figure, we can see that our network has a superior performance for both input meshes, even though our network was trained only on Gaussian noisy meshes.
Please refer to supplementary material Part 6 for more results on noise patterns.

Noting that Gaussian noise, uniform noise, and impulsive noise are all synthetic additive noises.  Hence, the differences between their distribution variations may not be very large (compared to real-scanned noise). 
As a result, our DNF-Net model trained only on Gaussian noise can still generalize well to handle other noise patterns.

\para{Robustness on noise intensities.} \
Lastly, we tested our network on 3D meshes corrupted with noise intensities larger than those in the training set.
Here, we used the training set of CNR~\cite{wang2016mesh} with Gaussian noise magnitude from $0.1\bar{l}_e$ to $0.3\bar{l}_e$, for which the released network of CNR was trained.
In this regard, for a fair comparison, we directly applied the trained models of both our network and CNR to denoise 3D meshes of noise magnitude above $0.3\bar{l}_e$, without re-training the networks.
Figure~\ref{fig:noise_level} plots the $\theta$ values for using various methods to denoise 3D meshes of increasing noise intensities, from $0.1\bar{l}_e$ to $0.7\bar{l}_e$.
From the plots, we can see that our method consistently achieves a better performance with the lowest $\theta$ values for all noise levels.
Please refer to supplementary material Part 7 for more results.

%%%%%%%%%%%%%%%%%%%%%%%%%%%%%%%%%%%%%%%%%%%%%%%%%%%%%%%%%%%%%%%%%%%%%

\subsection{Discussions}

\para{Time performance.} \
Our network takes only 0.04 seconds to process a patch on an NVidia Titan Xp GPU.
Further, since patch processing can be parallelized, our method's running time does not increase linearly with the number of patches.
Hence, to process thousands of patches on a dense mesh, e.g., a model with 50K faces, our method takes only 40 seconds.
If more GPUs are available, the computation time can be further shortened.

\para{Limitations.} \
First, as a common drawback of data-driven methods like~\cite{wang2016mesh}, our DNF-Net may produce unsatisfying denoising results, if we apply a network to process a test mesh whose noise pattern is very different from that in the training set.
We plan to explore domain adaptation techniques to extract or transfer knowledge from the unpaired data source.
Also, our method cannot handle meshes with faulty topological issues,~\eg, self-intersection and inconsistent facet normal orientation.
Such cases are also hard for existing filter-based and optimization-based methods.
Lastly, our network requires paired data (noisy meshes with ground truths) to train.
However, collecting a large amount of paired datasets is expensive and time-consuming.
In the future, we plan to explore the possibility of learning from unpaired data or training in an unsupervised manner.

%%%%%%%%%%%%%%%%%%%%%%%%%%%%%%%%%%%%%%%%%%%%%%%%%%%%%%%%%%%%%%%%%%%%%%%%%%%%%%%%%%%%%%%
\section{Conclusion}
\label{sec::conclusion}

This paper presents a novel deep normal filtering network, namely DNF-Net, formulated for mesh denoising.
DNF-Net is an end-to-end network that directly predicts denoised facet normals from noisy input meshes, without requiring explicit information about the underlying surface or the noise characteristics.
To effectively learn the local geometric patterns for denoising meshes, DNF-Net processes normal data grouped by patches.
Further, we design the multi-scale feature embedding unit to extract normal features, followed by the cascaded residual learning units to progressively remove noise.
Also, we drive DNF-Net to learn by formulating a deeply-supervised joint loss function with a normal recovery loss and a residual regularization loss.
Lastly, we performed several experiments on our methods using a rich variety of synthetic and real-scanned models.
Both visual and quantitative comparisons demonstrate the superiority of our method over the state-of-the-arts.

As the first attempt to design a deep neural network to filter facet normal for mesh denoising, our DNF-Net can yet be improved in several aspects.
First, instead of using only the normal data for denoising, we might as well consume other mesh information to enhance the extracted features,~\eg, vertex position, facet centroid, etc.
Second, we plan to explore graph convolutional networks to take into account the mesh topology in the network learning.
Third, enhancing the vertex update technique,~\eg, to handle local fold overs, will certainly help to improve the robustness of the overall method.
On the other hand, exploring techniques in domain adaptation and transfer learning is another future direction for improving the network generalization ability, particularly for handling real-scanned inputs.

\section*{Acknowledgments}
We thank anonymous reviewers for their valuable comments.
The work is supported by the Hong Kong Research Grants Council with Project No. CUHK 14225616, Key-Area Research and Development Program of Guangdong Province, China (2020B010165004), and National Natural Science Foundation of China with Project No. U1813204.
The work is also supported by the Research Grants Council of the Hong Kong Special Administrative Region (Project No. CUHK 14201717) and National Natural Science Foundation of China (Grant No. 61902275).

\bibliographystyle{IEEEtran}
\bibliography{egbib}
\if 0

\fi
\begin{IEEEbiography}[{\includegraphics[width=1in,height=1.25in,clip,keepaspectratio]{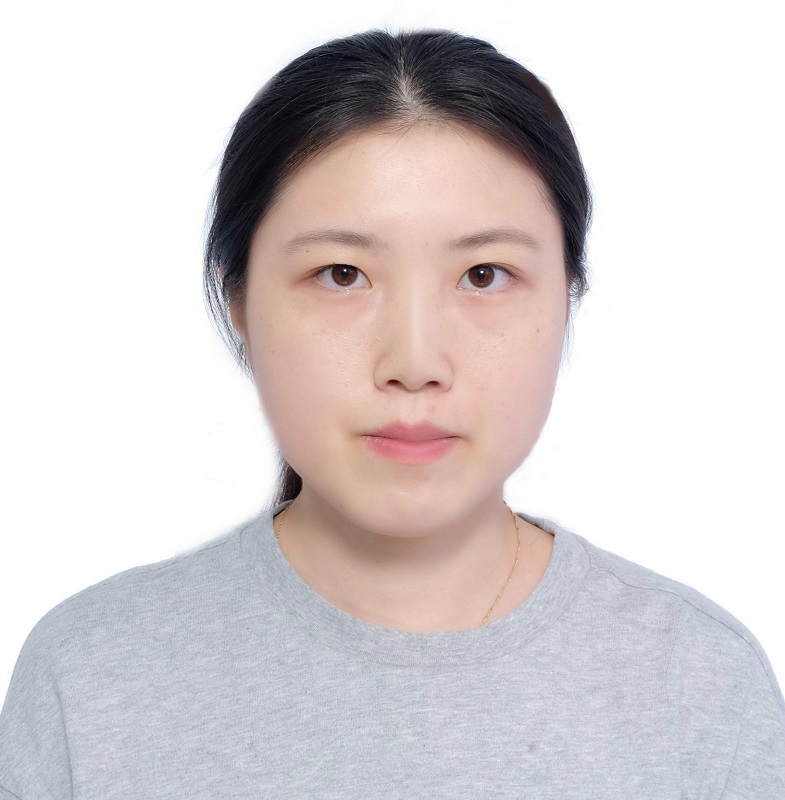}}]{Xianzhi Li}
	is currently a PhD student in the Department of Computer Science and Engineering, the Chinese University of Hong Kong. She will receive her Ph.D. degree in July, 2020. She served as the reviewers of CVPR 2020 and WACV 2021. Her research interests focus on geometry processing, computer graphics, and deep learning.
\end{IEEEbiography}
\vspace{-10mm}
\begin{IEEEbiography}[{\includegraphics[width=1in,height=1.25in,clip,keepaspectratio]{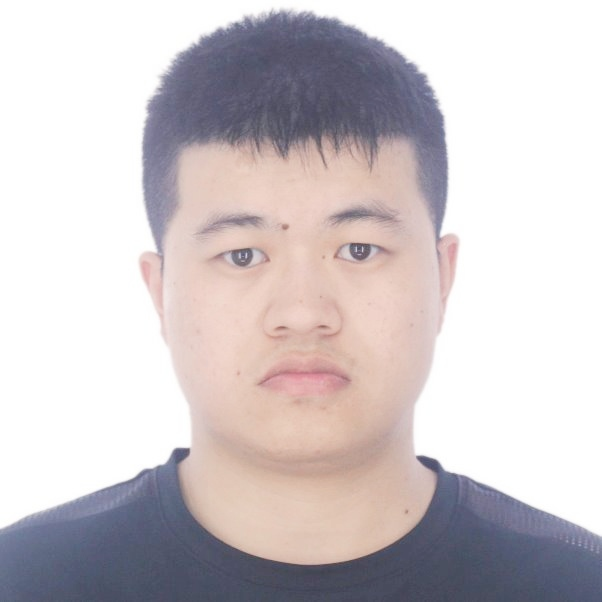}}]{Ruihui Li}
	is currently a PhD student in the Department of Computer Science and Engineering, the Chinese University of Hong Kong. His research interests include 3D processing, computer graphics, and deep learning.
\end{IEEEbiography}
\vspace{-10mm}
\begin{IEEEbiography}[{\includegraphics[width=1in,height=1.25in,clip,keepaspectratio]{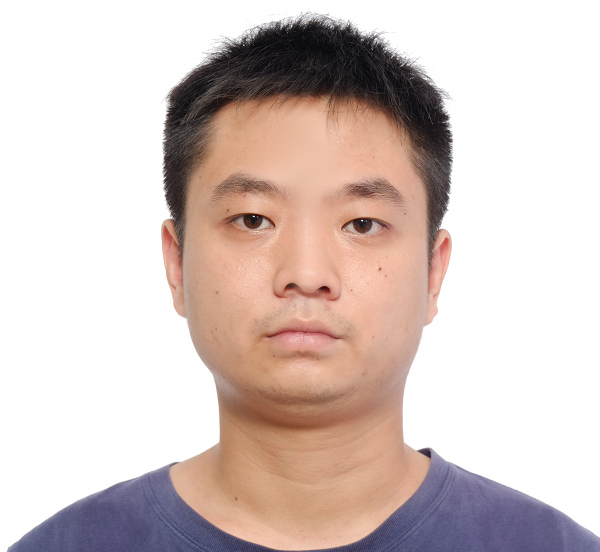}}]{Lei Zhu}
	received his Ph.D. degree in the Department of Computer Science and Engineering from the Chinese University of Hong Kong in 2017.
	He is working as a postdoctoral fellow in the Chinese University of Hong Kong. His research interests include computer graphics, computer vision, medical image processing, and deep learning.
\end{IEEEbiography}
\vspace{-10mm}
\begin{IEEEbiography}[{\includegraphics[width=1.0in,height=1.25in,clip,keepaspectratio]{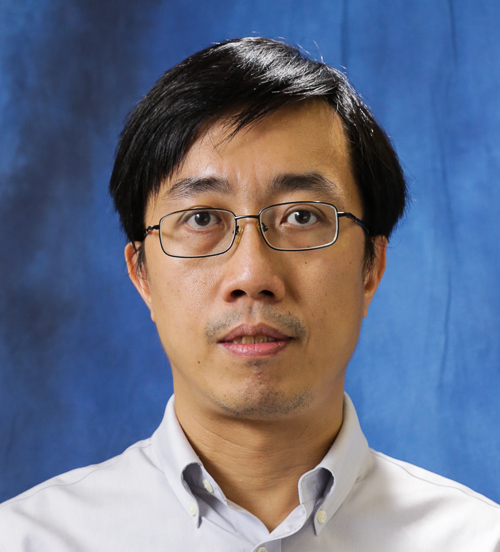}}]{Chi-Wing Fu} is currently an associate professor in the Chinese University of Hong Kong.  He served as the co-chair of SIGGRAPH ASIA 2016's Technical Brief and Poster program, associate editor of IEEE Computer Graphics \& Applications and Computer Graphics Forum, panel member in SIGGRAPH 2019 Doctoral Consortium, and program committee members in various research conferences, including SIGGRAPH Asia Technical Brief, SIGGRAPH Asia Emerging tech., IEEE visualization, CVPR, IEEE VR, VRST, Pacific Graphics, GMP, etc.  His recent research interests include computation fabrication, point cloud processing, 3D computer vision, user interaction, and data visualization.
\end{IEEEbiography}
\vspace{-10mm}

\begin{IEEEbiography}[{\includegraphics[width=1in,height=1.25in,clip,keepaspectratio]{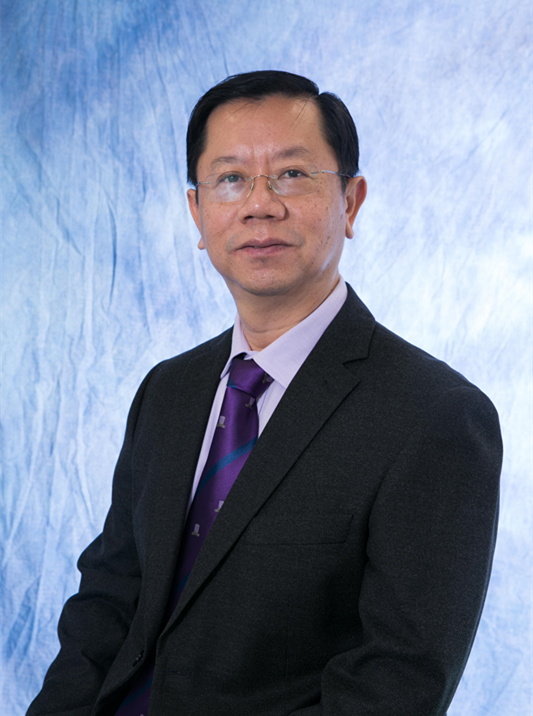}}]{Pheng-Ann Heng} received his B.Sc. from the National University of Singapore in 1985. He received his MSc (Comp. Science), M. Art (Applied Math) and Ph. D (Comp. Science) all from the Indiana University of USA in 1987, 1988, 1992 respectively. He is a professor at the Department of Computer Science and Engineering at The Chinese University of Hong Kong (CUHK). He has served as the Director of Virtual Reality, Visualization and Imaging Research Center at CUHK since 1999 and as the Director of Center for Human-Computer Interaction at Shenzhen Institute of Advanced Integration Technology, Chinese Academy of Science/CUHK since 2006. He has been appointed as a visiting professor at the Institute of Computing Technology, Chinese Academy of Sciences as well as a Cheung Kong Scholar Chair Professor by Ministry of Education and University of Electronic Science and Technology of China in 2007. His research interests include AI and VR for medical applications, surgical simulation, visualization, graphics and human-computer interaction.
\end{IEEEbiography}


%\newpage



%\vfill


\end{document}